\documentstyle[preprint,prb,aps]{revtex}

\begin{document}

\draft

\title{
Momentum Distributions in $^{\bf 3}$He-$^{\bf 4}$He Liquid Mixtures}

\author{J. Boronat,$^\dagger$ A. Polls$^\ddagger$ and A. Fabrocini$^\ast$}
\address{$^\dagger$ Departament de
F\'{\i}sica i Enginyeria Nuclear, Campus Nord B4-B5, \protect\\
Universitat Polit\`ecnica de Catalunya, E-08034 Barcelona,
Spain \protect\\
$^\ddagger$ Departament d'Estructura i Constituents de la Mat\`eria, \protect\\
Universitat de Barcelona, Diagonal 647, E-08028 Barcelona, Spain
\protect\\
$^\ast$ Istituto Nazionale di Fisica Nucleare, \protect\\
Dipartimento di Fisica, Universit\`a di Pisa,
Piazza Torricelli 2, I-56100 Pisa, Italy }

\maketitle

\date{\today}

\begin{abstract}

 We present variational calculations of the one-body density matrices and
momentum distributions for $^3$He-$^4$He mixtures in the zero temperature
limit, in the framework of the correlated basis functions theory.
The ground-state wave function contains two- and three-body correlations and
the matrix elements  are computed by (Fermi)Hypernetted Chain techniques.
The dependence on the $^3$He concentration $(x_3)$ of the $^4$He
condensate fraction $(n_0^{(4)})$ and of the  $^3$He pole
strength $(Z_F)$ is studied along the $P=0$ isobar.
At low $^3$He concentration, the computed  $^4$He condensate fraction
is not significantly affected by the $^3$He statistics. Despite of
the low $x_3$ values, $Z_F$ is found to be quite smaller
than that of the corresponding pure $^3$He  because  of the strong
$^3$He-$^4$He correlations and of the overall, large total density $\rho$.
A small increase of $n_0^{(4)}$ along $x_3$ is found, which is mainly
due to the decrease of $\rho$ respect to the pure $^4$He phase.

\end{abstract}

\pacs{  }

\narrowtext

The momentum distributions (MD) of atoms in quantum liquids is a challenging
problem of fundamental interest.\cite{siso89,glyde94} They provide
essential information on the correlations present in the system, which do
not show up explicitly in other quantities. In the past years, 
accurate inelastic
neutron scattering experiments have allowed for studying  several aspects
of the momentum distribution in helium liquids, $^4$He,\cite{sok89,sok92}
 $^3$He \cite{sok85} and $^4$He-$^3$He mixtures.\cite{sok94,sok95}
However, a clean extraction of information on the Helium MD's is
 somehow tampered by the need of a sound theoretical understanding of the
final state effects in the analysis of the dynamic structure function, even
at high momentum transfers.

The theoretical methods to evaluate momentum distributions of
many-body interacting, dense systems at zero temperature
have also made a significant progress in recent
years.\cite{siso89} At present, there are  results for the
pure Helium phases obtained within different many-body techniques,
i.e., variational theory (using either integral equations 
\cite{manou85,fabro92} or
Monte Carlo methods \cite{VMC}) and almost exact stochastic methods
as Green's Function Monte Carlo (GFMC) \cite{pa87,boro94} or
Path-Integral Monte Carlo (PIMC ).\cite{ceper92}

The MD's of liquid $^4$He ($^3$He) are
influenced by the Bose (Fermi) statistics of the
atoms. The macroscopic occupation of the zero momentum state, as given
by the condensate fraction $n_0^{(4)}$, characterizes the momentum
distribution of bosonic, liquid $^4$He and it is strictly linked to 
its superfluid behavior. On the
other hand, the discontinuity $Z_F$ at the Fermi momentum $k_F$ is a
characteristic of the $^3$He system when it is studied as a normal
Fermi liquid.

In this paper we consider the interesting case of isotopic $^3$He-$^4$He
mixtures where, due to its fermion -boson nature, both quantities $Z_F$ and
$ n_0^{(4)}$ are simultaneously present.  Recent neutron scattering
experiments on Helium mixtures at high momentum
transfers \cite{sok94,sok95} give  additional motivations to undertake a
microscopic, theoretical
study of their momentum distributions and one-body density matrices.
Special emphasis will be devoted to the dependence on the $^3$He
concentration, $x_3$, of the single-particle kinetic energies of the
isotopes and of $Z_F$ and $n_0^{(4)}$.

The investigation is carried on in the framework of the variational
approach. The trial wave function for the mixture contains two--body
(Jastrow) and triplet correlations. This type of correlated wave function
has been useful in effectively studying the pure phases.
\cite{manou85,fabro92,fanto82,manou83} Two of us \cite{fabro182}
 (A.P. and A.F.) derived the hypernetted and Fermi
hypernetted chain (HNC/FHNC) equations for the momentum distributions of
the mixtures using trial wave functions with
only pair correlations. Numerical applications were carried out in the
HNC/FHNC/0 approximation, i.e., neglecting the elementary
diagrams.
A preliminary study of the elementary diagrams for a Jastrow trial wave
function was performed \cite{boro90} by generalizing the scaling approximation
proposed for pure phases.\cite{manou85,fabro92} Also available are variational
Monte Carlo (VMC) calculations \cite{lee81} with similar correlations
of the analytical McMillan type.
The studies of the mixture has been recently complemented
with variational calculations concerning the energy and stability of the
ground state,
\cite{kro93,boro93} with path integral Monte Carlo (PIMC) analysis
\cite{boni95} and with microscopic correlated basis functions (CBF) estimates
of the inelastic neutron scattering cross sections both at intermediate
\cite{fabro96} and high \cite{mazzanti95} momentum transfers.

The paper is organized as follows: in the second section we will shortly
present the HNC/FHNC theory to calculate $n(k)$ for mixtures described
by correlated wave functions containing two- and three-body
correlations. The treatment of the elementary diagrams in the so called
scaling approximation is discussed in some details in the second part
of the section.
Results for $n^{(4)}(k)$, $n^{(3)}(k)$ and for the one-body density matrices
are presented in Section II, together with a critical discussion of
the discrepancies with the available analysis of the deep inelastic
neutron scattering measurements on mixtures, which (in contrast with
our results) point to a large enhancement of the $^4$He condensate fraction.

\section{
 HNC/FHNC Equations for the Momentum Distribution of
 $^{\bf 3}$He-$^{\bf 4}$He mixtures}

The one-body density matrices $\rho^{(\alpha)}({\bf r}_1,{\bf
r}^{\prime}_1)$
($\alpha=3,4$) for a homogeneous, isotopic mixture of $N_3$ $^3$He atoms 
and $N_4$
$^4$He atoms, described by a ground-state wave function
$\Psi(1,\ldots,N_4+N_3)$ are defined as

\begin{equation}
\rho^{(\alpha)}({\bf r}_1,{\bf r}_1^{\prime})=\frac{N_\alpha}{\rho_\alpha}
\frac{\int \Psi^*(1_\alpha,\ldots,N_4+N_3) \Psi(1^{\prime}_\alpha,\ldots,
N_4+N_3) d{\bf r}_2  \ldots d{\bf r}_{N_4+N_3}}{\int
\mid \Psi (1,\ldots,N_4+N_3)\mid ^2 d{\bf r}_1 \ldots d{\bf
r}_{N_4+N_3}} \ .
\label{onebody}
\end{equation}

In homogeneous mixtures, with constant particle densities
$\rho_{\alpha}=N_\alpha/N$,
$\rho^{(\alpha)}({\bf r}_1,{\bf r}_1^{\prime})=\rho^{(\alpha)}(r)$, with
$r=\mid {\bf r}_1-
{\bf r}_1^{\prime} \mid$. $\rho^{(\alpha)}(r)$'s satisfy the normalization
conditions
$\nu_{\alpha} \rho^{(\alpha)}(0)=1$, $\nu_{\alpha}$ being the spin
degeneracy ($\nu_4=1$, $\nu_3=2$). Notice that in the definition of
$\rho^{(3)}(r)$ the spin variables have not been explicitly written.
We will henceforth omit  the subindex in the degeneracy factor and assume that
it always refers to $^3$He.

The momentum distribution of the $\alpha$ component, or rather the
occupation
probability for single-particle states with momentum ${\bf k}$ and given
spin projection, can be obtained as the Fourier transform of
the corresponding density matrix,
\begin{equation}
n^{(\alpha)}(k)=\delta_{\alpha 4} \rho_4 n_0^{(4)} (2 \pi)^3
\delta({\bf k})+ \rho_{\alpha}
 \int d{\bf r} \exp(i {\bf k}\cdot {\bf r})
[\rho^{(\alpha)}(r)-\delta_{\alpha 4} n_0^{(4)}] \ ,
\label{enek}
\end{equation}
where $n_0^{(4)}=\rho^{(4)}(\infty)$ is the $^4$He condensate fraction,
i.e., the fraction of $^4$He particles in the zero momentum
state.

The ground state of the mixture is well described by a generalization of the
correlated wave function
 used in the pure phases:
\begin{equation}
\Psi(1,\ldots,N_4+N_3)= \prod _{\alpha \leq \beta \leq \gamma
=3,4}  \prod _{i_{\alpha}\leq j_{\beta}}f^{(\alpha,\beta)}(i_\alpha,j_\beta)
 \prod _{i_{\alpha}\leq j_{\beta} \leq k_{\gamma}} f^{(\alpha,\beta,\gamma)}
(i_\alpha,j_\beta,k_\gamma) \  \phi(1,\ldots,N_3).
\label{eq:fona}
\end{equation}
 $\phi(1,\ldots,N_3)$ is the Slater determinant of
plane
waves corresponding to the Fermi component of the mixture, and 
$f^{(\alpha,\beta) }(i_{\alpha},j_{\beta}$) (
$f^{(\alpha,\beta,\gamma)}(i_{\alpha}, j_{\beta},
k_{\gamma})$) are the 2 (3)-body correlation functions involving 2 (3)
 particles
of types $\alpha,\beta$ ($\alpha,\beta,\gamma$), respectively.
Similar trial wave
functions have been used in previous works to study the structure and energetic
ground-state properties of  $^3$He-$^4$He mixtures.
\cite{fabro182,kro93,boro93}

A cluster analysis of $\rho^{(\alpha)}(r)$ in powers of
 $\omega^{(\alpha,\beta)}\equiv f^{(\alpha,\beta)}-1$,
$h^{(\alpha,\beta)}\equiv
[f^{(\alpha,\beta)}]^2-1$, $\omega^{(\alpha,\beta,\gamma)}\equiv
f^{(\alpha,\beta,
\gamma)}-1$ and $h^{(\alpha,\beta,\gamma)}\equiv
[f^{(\alpha,\beta,\gamma)}]^2-1$,
as that carried out in  the pure phases,
\cite{ris76,fanto78}
gives the following structural decomposition for $\rho^{(\alpha)}(r)$:
\begin{equation}
\rho^{(\alpha)}(r)= n_0^{(\alpha)} N^{(\alpha)}(r) \ ,
\label{ralfa}
\end{equation}
where massive re-summations of the diagrams, 
as defined in Refs. \onlinecite{manou85,fabro92,fabro182,fanto78},
may be performed in practice by using HNC/FHNC
techniques.\cite {fabro182,boro93,fabro82}

The strength factor $n_0^{(\alpha)}$ is given by
\begin{equation}
n_0^{(\alpha)} = \exp[2 \Gamma_{\omega}^{(\alpha)}-\Gamma_d^{(\alpha)}]
\label{noalfa}
\end{equation}
and
\begin{equation}
N^{\alpha}(r)=[\delta_{\alpha 4}+\delta_{\alpha 3}(\frac{1}{\nu}
l(
k_Fr)- N_{\omega_c \omega_c}^{(3)}(r)-E_{\omega_c \omega_c}^{(3)}(r))]
\exp{[N_{\omega \omega}^{(\alpha)}(r)+E_{\omega \omega}^{(\alpha)}(r)]}
\
\label{nalfa}
\end{equation}
 sums up all the irreducible diagrams with external points
$1_{\alpha}$ and $1_{\alpha}^{\prime}$. In Eq.(\ref{nalfa}), $l(x)=3
j_1(x)/x$ is the Slater function and  $k_F=(6\pi ^2 \rho/\nu)^{1/3}$
is the $^3$He Fermi momentum.

The functions $N^{(\alpha)}_{xy}(r)$ and $E^{(\alpha)}_{xy}(r)$ are 
the sums of the {\em nodal} and {\em elementary} diagrams contributions, 
respectively. The evaluation of the nodal functions $N^{(\alpha)}_{xy}(r)$, 
in the context of the HNC/FHNC approach, 
is discussed in the Appendix, also containing the explicit expressions 
of the $\Gamma_{\omega,d}^{(\alpha)}$ factors. 

The momentum distributions are computed via the density matrices by 
Eq.(\ref{enek}). We thus get

\begin{equation}
n^{(4)}(k)=(2 \pi)^3 \rho_4 n_0^{(4)} \delta({\bf k})+\rho_4 n_0^{(4)}
\int d{\bf r}
\exp [i{\bf k}\cdot {\bf r}] (\exp [N_{\omega \omega}^{(4)}(r)+E_{\omega
\omega}^{(4)}(r)]-1)   \ ,
\label{n4fin}
\end{equation}
and
\begin{equation}
n^{(3)}(k)=n_0^{(3)}[n_{c}(k)+\Theta(k_F-k) n_{d}(k)] \ ,
\label{n3fin}
\end{equation}
where
\begin{equation}
n_{d}(k)=1-\tilde X_{cc}+2 \tilde
X_{\omega_cc}+\frac{\tilde X_{\omega_cc}^2}
{1-\tilde X_{cc}}
\label{ndis}
\end{equation}
and
\begin{eqnarray}
n_{c}(k) & = & -\frac{\tilde X_{\omega_cc}^2}{1-\tilde X_{cc}}-\rho_3
\int d{\bf r}
\exp[i{\bf k}\cdot{\bf r}] \left \{(\exp[N_{\omega
\omega}^{(3)}(r)+E_{\omega \omega}^{(3)}(r)] -1)     \right .   \\
\nonumber & &  \left . \times (-l(k_Fr)/\nu + N_{\omega_c
\omega_c}^{(3)}(r)+ E_{\omega_c \omega_c}^{(3)}(r))+
E_{\omega_c \omega_c}^{(3)}(r)\right \} \ .
\label{ncon}
\end{eqnarray}
 $X_{yc}=g_{yc}-N_{yc}+l/\nu$ for $y=\omega_c, c$ and
$\tilde X_{xy}(k)$ stands for the Fourier transform
\begin{equation}
\tilde X_{xy}(k)=\rho_3
 \int d{\bf r} \, e^{i {\bf k}\cdot {\bf r}} X_{xy}(r)
\label{xalal}
\end{equation}

The strength factor $n_0^{(4)}$ is the asymptotic value of the $^4$He 
one-body density
matrix, $ \rho^{(4)}(r\rightarrow \infty) = n_0^{(4)}$ and
corresponds to the $^4$He condensate fraction.
The decomposition of $n^{(3)}(k)$ in a continuous
($n_{c}(k)$) and a discontinuous ($n_{d}(k)$) piece explicitly links
the discontinuity of $n^{(3)}(k)$ at $k_F$, $Z_F$, to $n_d(k_F)$ by
\begin{equation}
Z_F=n_0^{(3)} n_d(k_F)  \ .
\label{eq:zz}
\end{equation}

\subsection{Scaling approximation for the elementary diagrams}

The HNC/FHNC equations can be solved once a given prescription
for the contributions of the elementary diagrams has been given.
However, as no exact method to compute them is presently known, 
at least in the frame of the integral equations, one has to resort to 
some approximation.
Among the available  schemes\cite{usma82,rosa82,oddi89} we have chosen the
scaling approximation (SA), developed for both the energy
and the one-body density matrix of pure phases,
\cite{manou85,fabro92,fanto82,manou83} and satisfactorily
reproducing VMC calculations.
Although the number of elementary diagrams in the mixture is much larger,
it is straightforward to generalize the pure phases scaling approximation
to our case.

The SA is based on the evaluation of the 4-points
elementary diagrams constructed with the combinations of the distribution
functions $g_{xy}^{(\alpha,\beta)}(r)$ allowed by the diagrammatic rules
and it has already been used in the  calculation of the energy and of the
static structure functions of the mixture.\cite{boro93}
The elementary diagrams are approximated by
\begin{equation}
E_{dd}^{(\alpha,\beta)}(r)=E(r)~~~, E_{xy}^{(\alpha,\beta)}(r)=0,~~~
\alpha,\beta \in \{3,4\}, \ xy =[de,ee,cc]  \ ,
\label{eq:ele}
\end{equation}
where
\begin{equation}
E(r)=(1+s) E_g^{[4]}(r)~+~E_t^{[4]}(r) \ .
\label{eq:scaling}
\end{equation}
$E_g^{[4]}(r)$ and $E_t^{[4]}(r)$ are the four-points elementary diagrams
without and with explicit three-body correlations into their  basic
structure, respectively. These diagrams are constructed by using as internal
links an $averaged$ dressed correlation $\hat g(r)-1$,

\begin{equation}
\hat g(r)= x_4^2 \, g^{(4,4)}(r) + 2 x_3 x_4 \, g^{(4,3)}(r) + x_3^2 \,
g^{(3,3)}(r)\ ,
\label{eq:gave}
\end{equation}
with $x_\alpha=\rho_\alpha/\rho$. The introduction of $\hat g(r)$ makes 
feasible the calculation of $E(r)$ because
it reduces drastically the high number of elementary diagrams originated
by all the possible bonds between $^3$He and $^4$He particles.
Actually, for the underlying boson-boson mixture (i.e. $\Phi(1,\ldots,N_3)=1$ 
in Eq. (\ref{eq:fona})) and taking the same
correlation functions between all types of isotopes ( average
correlation approximation (ACA) ),  $\hat g(r)$ provides the
exact $E_{g,t}^{[4]}(r)$. This property
and the small $^3$He concentration in the physical region
of interest ($x_3 < 0.10$) justify the use of $\hat g(r)$. The
scaling parameter $s$ (\ref{eq:scaling}) is determined by imposing
the consistency between
the Pandharipande-Bethe and the Jackson-Feenberg forms of the kinetic
energy for the boson-boson mixture without triplet  correlations.
$s$ is calculated for each total density and it is kept fixed when 
$x_3$ changes.
This assumption is plausible because, at low $^3$He concentrations, the
statistical effects in $\hat g(r)$ are negligible.

The additional elementary diagrams needed for  the one-body density
 matrices are similarly evaluated:
\begin{equation}
E_{\omega d}^{(\alpha,\beta)}(r)= E_{\omega d}(r),~~~~~
E_{yz}^{(\alpha,\beta)}=0 ~~~(yz=\omega e, \omega_c c)
\end{equation}
with
\begin{equation}
E_{\omega d}(r)=(1+s_{\omega d}) E_{\omega d,g}^{[4]}(r)+E_{\omega
d,t}^{[4]}(r),
\end{equation}
and
\begin{equation}
E_{\omega \omega}^{(\alpha)}(r)=(1+s_{\omega \omega}^{(\alpha)})
E_{\omega \omega,g}^{[4]} + E_{\omega \omega,t}^{[4]}(r),
\end{equation}
\begin{equation}
E_{\omega_c \omega_c}^{(3)}(r)=(1+s_{\omega_c \omega_c}) E_{\omega_c
\omega_c,g}^{[4]}+ E_{\omega_c \omega_c,t}^{[4]}(r).
\end{equation}
The average distribution function
\begin{equation}
\hat g_{\omega}(r)= x_4^2 \, g_{\omega d}^{(4,4)}(r) + 2 x_3 x_4
\, (g_{\omega d}^{(4,3)}(r) + g_{\omega e}^{(4,3)}(r))+ x_3^2 \, (g_{\omega
d}^{(3,3)}(r) + g_{\omega e}^{(3,3)}(r))
\end{equation}
has been used to compute the above four-points elementary diagrams.

Finally, the set of single external point elementary diagrams, appearing in the
strength factors $n_0^{(\alpha)}$ expressions, are approximated, as in the
pure phases,\cite{manou85,fabro92} by
\begin{equation}
E_x= (1 + {3 \over 2}s_{xd}) E_{x,g}^{[4]}+ E_{x,t}^{[4]} \ , ~~~~
x=\omega,d~.
\end{equation}

As far as the factors related to the momentum
distributions are concerned, we have chosen
$s_{\omega d}$ by imposing $T_{MD}=T_{JF}$, where  $T_{MD}$ is
the total kinetic energy obtained by integrating the momentum
distribution,
\begin{equation}
T_{MD}= {{\hbar^2}\over {2 m_4}}{x_4 \over {(2 \pi)^3 \rho_4}} \int
d {\bf k}~ k^2~ n^{(4)}(k) + {{\hbar^2}\over {2 m_3}} {{x_3 \nu}\over {(2
\pi)^3
\rho_3}} \int d{\bf k}~ k^2~ n^{(3)}(k),
\label{tmd}
\end{equation}
 and $T_{JF}$ is the ground-state expectation value of the kinetic
energy operator computed by the Jackson-Feenberg identity.
Moreover, the fulfillment of the normalization conditions of the momentum
distributions, i.e.,
\begin{equation}
{{\nu_{\alpha}}\over {(2\pi)^3 \rho_{\alpha}}} \int d{\bf k} \, 
n_{\alpha}(k)~=~1~,
\end{equation}
equivalent to $ \nu_{\alpha} \rho^{(\alpha)}(0)=1$, requires
\begin{equation}
n_0^{(\alpha)} \exp \left [{N_{\omega \omega}^{(\alpha)}(0) + E_{\omega
\omega}^{(\alpha)}(0)}\right ]= ~1
\label{renor1}
\end{equation}
\begin{equation}
N_{\omega_c \omega_c}^{(3)}(0) + E_{\omega_c \omega_c}^{(3)}(0) =  0 \ .
\label{renor2}
\end{equation}
These conditions are used to determine the remaining scaling parameters
($s_{\omega \omega}^{(\alpha)}$, $s_{\omega_c \omega_c}$).

As a matter of fact, the use for the triplet correlated wave function 
of the same $s_{\omega \omega}^{(\alpha)}$ and $s_{\omega_c \omega_c}$ 
parameters, as determined in the Jastrow case, produces significant deviations 
of the above normalizations from their exact values.
For this reason and to ensure the correct normalizations of the density 
matrices, we have recalculated the scaling factors
$s_{\omega d}$, $s_{\omega \omega}^{(4)}$, $s_{\omega \omega}^{(3)}$
and $s_{\omega_c \omega_c}$ when the wave function contains three-body
correlations, as in Ref. \onlinecite{fabro92}.

\section {Results}

In this section we report results for the momentum distributions of
$^3$He-$^4$He liquid
mixtures using the Aziz potential (HFDHE2)\cite{aziz} for the variational
determination of the ground-state correlations. This interaction effectively 
describes the equation of state of the pure phases.  \cite{boro94,pa89} 
The interatomic potential in isotopic mixtures is the
same between any pair of particles. Based on this fact, we have used
the average correlation approximation, ACA. The ACA approach, which has
been carefully analyzed for the impurity problem,\cite{boro89} has also
been used in the past to study  finite concentration Helium mixtures.
\cite{boro93,fabro84,guyer80} The potential is strongly repulsive at
short distances, so the correlation functions are expected to show the
same short-range behaviors. Small differences
can arise however at intermediate and large distances, where the interaction is 
weaker,
because of the different  masses and statistics of the isotopes.
Nevertheless, ACA may well serve to the purpose of studying the
$x_3$-dependence  of the momentum distributions in the mixture.
In fact, for Jastrow correlated wave functions we have released the ACA,
allowing for different correlations in different isotopic pairs, and  these
extra variational degrees of freedom have  not significantly
changed our results.

The two-body correlation
function $f(r)$ has been taken to have an analytical form,
of the McMillan type at short distance and with enough
flexibility to adjust to the optimal pure $^4$He
correlation behavior in the intermediate and long ranges,

\begin{equation}
f(r)= \exp {\left(-{1 \over 2} \left( {{b }\over
{r}}\right)^5
\right)} \left[ A + B \exp { \left( - {{(r - D)^2} \over {\tau
r^4}}\right)} \right ].
\label{f2}
\end{equation}
The  long range, $r^{-2}$ behavior ensures the proper linear
dependence of the $^4$He structure function at $k \rightarrow 0$.

The $f(r)$ parameters at the $^4$He energy variational minimum,
at equilibrium density $\rho_0 =0.365 \ \sigma^{-3}$ ($\sigma = 2.556$
\AA), are
$b=1.18\ \sigma$, $~A= 0.85,~B=1-A,~D=3.8$ \AA~ and $\tau=0.043 $ \AA$^{-2}$.
$B$ and $\tau$ are related to the experimental pure
$^4$He sound velocity $c$  and to the low-$k$ behavior of its static
structure function by
\begin{equation}
{B \over \tau}= {{m_4 c}\over {2 \pi^2 \hbar \rho_0}} ~.
\end{equation}

The three-body correlation function $f(r_{ij},r_{ik},r_{jk})$ has the
parameterized form:\cite{manou85,fabro92,fanto82,manou83}
\begin{equation}
f(r_{ij},r_{ik},r_{jk}) = \exp {\left [- {1 \over 2} \sum_{l=0,1} \lambda_l
  \sum_{cyc} \xi_l(r_{ij}) \xi_l(r_{ik})
P_l({\hat r}_{ij}\cdot {\hat r}_{ik}) \right ] },
\label{f3}
\end{equation}
where
\begin{equation}
\xi_l(r)= (r -\delta_{l0}r_{tl}) \exp { \left [ - \left ({{r -r_{tl}}\over
{\omega_{tl}}} \right)^2 \right ] }.
\end{equation}
The values of the triplet functions parameters have been taken from
Ref. \onlinecite{fanto82} omitting the small $l=2$ component.

The calculations presented here are performed at the experimental values
of the density along the $P=0$ isobar. In this regime, the density decreases
from $\rho=\rho_0$ ($x_3=0$) to $\rho=0.3582\ \sigma^{-3}$ at $x_3=0.066$,
corresponding to the $^3$He maximum solubility.
The partial $^3$He density increases from zero up to $\rho_3=0.0236\
\sigma^{-3}$
in the same $x_3$ range. So, we have neglected the density
dependence of the variational parameters of the correlations
because of the small variations both of the total and partial densities
 in the region of physical interest.

Before presenting the results for the Helium mixtures, it is worthwhile to
study the accuracy of the scaling approximation in the case of pure
$^4$He. We have considered a correlated wave function containing 
McMillan two-body correlations ($A=1.$, $B=0.$, and
$b=1.20\ \sigma$ in Eq.(\ref{f2})) and a three-body factor given 
by Eq.(\ref{f3}). 
At ${\rho_0}$ we obtain
$n_0^{(4)}({\rm JT}_1)=0.078$ and $n_0^{(4)}({\rm JT}_{01})=0.081$,
where the JT$_{1}$ (JT$_{01}$) results include triplet correlations 
contributions without (with) the $l=0$ component. 
The corresponding energies are $E/N ({\rm JT}_1)=-6.55$ K
and $E/N ({\rm JT}_{01})=-6.58$ K. 
A VMC study by one of the authors (J.B.), with the same trial wave 
functions, gives  
$n_0^{(4)}({\rm JT}_1)({\rm VMC})=0.078$, $n_0^{(4)}({\rm JT}_{01})({\rm
VMC})=0.082$,  
$E/N ({\rm JT}_1)({\rm VMC})=-6.617$ K and
$E/N ({\rm JT}_{01})({\rm VMC})=-6.625$ K. These results have been 
confirmed by an independent VMC calculation of Moroni,\cite{moropriv} who  
gets $n_0^{(4)}=0.077$ and  $E/N =-6.604$ K for the (JT$_1$) case.

The agreement between HNC and VMC results gives confidence in the
scaling approximation to the elementary diagrams as
described in the previous section, prescribing a recalculation of
the scaling parameters directly associated to the momentum distribution
after the inclusion of the three-body correlations. Actually, if the scaling
parameters in the JT cases are the ones determined at the Jastrow level (as
in Refs. \onlinecite{manou85,manou91}), we get $n_0^{(4)}({\rm JT}_1)=0.064$ 
with a violation of the normalization conditions of $\sim 15 \%$.
In addition, the $l=0$ component of the triplet correlation has been found 
to have a very small effect on both the energy and condensate fraction. 
This finding also has been confirmed by the Moroni calculations\cite{moropriv}
and is in contrast with that of Refs. \onlinecite{manou85,manou91}, 
 where the relative change  in $n_0$ was about 25 \%. 
Due to the small effect of the $l=0$ triplet
correlation, we have omitted its contribution in all the results
presented for the mixture.

The use of the semi-optimized two-body correlation factor of Eq.(\ref{f2}) 
and of the $l=1$ triplet correlation lowers the energy to
$-6.62$ K and provides $n_0^{(4)}=0.082$.
The Euler Monte Carlo (EMC) result of Ref. \onlinecite{moropriv}, 
using fully  optimized two- and three-body correlations in a VMC scheme,
is $n_0^{(4)}({\rm EMC})=0.087$.
On the other hand, the DMC results of Refs. \onlinecite{moro97,boro94}
are $n_0^{(4)}({\rm DMC})= 0.072$ and $n_0^{(4)}({\rm DMC})= 0.084$, 
respectively.
The difference between the two DMC results is due to the
use of an extrapolated estimator which is
 sensitive to  the overlap between the importance sampling wave function and 
the exact ground state.
The PIMC approach of Ref. \onlinecite{ceper92} provides 
$n_0^{(4)}({\rm PIMC})=0.069$ at temperature T=1.18 K, with large statistical 
errors.
 As a final comment, we stress that all the above theoretical values  
of the $^4$He condensate fraction are slightly lower than the latest 
experimental estimates of Snow {\it et al.},\cite{snow92} 
$n_0^{(4)}({\rm expt})\sim 0.10$. However, as the condensate fraction, 
as well as 
the kinetic energy, is extracted by fitting the Compton scattering profile in 
neutron scattering experiments at large momentum transfers, the resulting 
$n_0^{(4)}$ can be strongly model dependent.

We start the analysis of the mixture by studying the $x_3$-dependence of
$^4$He momentum distribution.
Fig. 1 shows $k n^{(4)}(k)/((2 \pi)^3 \rho_4)$   in  mixture at
$x_3=0.066$
($\rho_{\rm expt}= 0.358 \ \sigma^{-3}$) compared with that of pure $^4$He 
($\rho_4= 0.365 \ \sigma^{-3}$),
both at P=0 . The differences are small and can be explained by the
slight change in density.
In fact, the smaller mass of
$^3$He results in a larger zero point motion of
$^3$He compared with $^4$He, and therefore the total density of the
mixture decreases when $x_3$ increases.

Fig. 2 illustrates the same comparison but for the $^4$He 
one-body density matrix. The asymptotic value of $\rho^{(4)}(r)$, 
identified with the condensate fraction,  is reached at $r \sim 7$  \AA. 
The value of $n_0^{(4)}$  in the mixture
is slightly larger than in the pure phase (see also Table I) due mainly to the
smaller total density
of the mixture. The fermionic nature of the $^3$He
does not affect $n_0^{(4)}$. In fact, one gets the
same $n_0^{(4)}$ in the boson-boson
approximation, which consists in treating the $^3$He component as a 
bosonic mass-3 one.
Furthermore, if ACA is assumed, the boson-boson approximation yields a
$n_0^{(4)}$ which
is exactly  the one of pure $^4$He at the total density of the mixture.

The Fermi statistics makes the $x_3$-dependence of
$n^{(3)}(k)$ more sizeable. The $^3$He momentum distributions at $x_3=0.066$ 
and $x_3=0.020$ are compared in Fig. 3. The corresponding Fermi momenta  
are $k_F=0.235$ \AA$^{-1}$ and $k_F=0.347$ \AA$^{-1}$,  
to be compared with $k_F= 0.79$ \AA$^{-1}$ for pure $^3$He
at equilibrium density. The Fermi momentum and the discontinuity $Z_F$ 
increase along $x_3$ , whereas the depletion decreases (see Table
I). This behavior is qualitatively explained by considering the change of 
both the total  and  partial $^3$He densities.

$\rho^{(3)}(r)$ at $x_3=0.066$ is compared in Fig. 4 with the free 
fermionic case 
($\nu \rho(r)/\rho = l(k_F r)$) and with that of pure
$^3$He at the same $\rho_3$. In this density region it 
is necessary to reach large $r$-values
before $\rho^{(3)}(r)$ begins to oscillate around zero.
 Despite of the small partial $^3$He density, $\rho^{(3)}(r)$ is very different
from those obtained both in the pure (short-dashed line) and 
the free (long-dashed line) cases. While the pure $^3$He shows a density
matrix very similar to the free case, the mixture $\rho^{(3)}(r)$ has a strong
depletion due to the correlations with the $^4$He atoms. This behavior 
translates into a correspondingly large depletion of
$n^{(3)}(k)$ at the origin. The three density matrices have the nodes 
approximately at the same points, the location of the zeros  being 
governed by the  zeros of $l(k_F r)$. In fact, by taking the lowest 
order term of the expansion of $\rho^{(3)}(r)$ in powers of the statistical 
correlation $l(k_Fr)$, as done in the Wu-Feenberg expansion
for the distribution function, one gets
\begin{equation}
\rho^{(3)}_{WF}(r) = \rho^{(3)}_B(r) {{l(k_F r)}\over {\nu}}
\label {eq:decoup}
\end{equation}
where $\rho^{(3)}_B(r)$ is the $^3$He density matrix in the
underlying boson-boson mixture. Due to the small values of $x_3$ in the mixture,
$\rho^{(3)}_{WF}(r)$ is almost indistinguishable 
from the exact $\rho^{(3)}(r)$.

Eq. (\ref{eq:decoup}) explicitly decouples the statistical and dynamical
correlations contributions to $\rho^{(3)}(r)$ and  has also recently proved  
to describe quite accurately even  the pure $^3$He density matrix.\cite{moro97}
In this approximation, $n^{(3)}(k)$ is given by
\begin{equation}
n^{(3)}_{WF}(k) = {{1}\over {(2 \pi)^3 \rho_3}} \int_0^{k_F} d^3 k' n_B^{(3)} 
(\mid {\bf k} - {\bf k}' \mid)   \ .
\label{eq:ndecoup}
\end{equation}
Therefore, the discontinuity $Z_F$ coincides with the value of the
condensate fraction associated to $n^{(3)}_B(k)$. The kinetic energy associated to
$n^{(3)}_{WF}(k)$ can be expressed as
\begin{equation}
\frac{T_3}{N_3} = \frac{3 \hbar^2 k_F^2}{10 m_3} + \frac{T_{B3}}{N_3} \ ,
\label{tsimple}
\end{equation}
where $T_{B3}/N_3$  is the kinetic energy associated to $n_B^{(3)}(k)$.
In ACA, the density matrices of the two components of the underlying
boson-boson mixture are the same and equal to the density matrix of
pure $^4$He considered at the
total density of the mixture. As a consequence, the corresponding
condensate fractions are also equal and in this model $Z_F$ and
$n_0^{(4)}$ coincide.

More detailed information on the $x_3$-dependence of the condensate
fraction,
the discontinuity of $n^{(3)}(k)$ at the Fermi surface and the kinetic
energies of the two components is shown in Table I. $T_3(x_3=0.)$ is 
the kinetic energy of one $^3$He impurity in $^4$He.  
Recent DMC\cite{boro96} and PIMC\cite{boni95} calculations
predict a smaller $T_3(x_3=0.)$ value of about 17.5 K.
The effect of the
three-body correlations is similar to that in the $^4$He pure phase, i.e.,
they slightly decrease the condensate fraction
and simultaneously decrease by about half a Kelvin the total kinetic energy.
The condensate fraction $n_0^{(4)}$ shows a small increment with $x_3$. 
As we have mentioned before, this is mainly a consequence of the fact
that the  total density of the mixture slightly decreases when $x_3$
increases. 
The effect of the Fermi statistics on $n_0^{(4)}$ is
almost negligible, the results of $n_0^{(4)}$ in the boson-boson
approximation being equal to the ones reported in Table I.

$n_0^{(4)}$ is shown in Fig. 5 as a function of the 
pressure, $P$, for pure $^4$He (diamonds) and for a $x_3=0.066$ mixture 
(circles).
The condensate fraction, in both cases, decreases with pressure as a consequence
of the corresponding increase of density. The density of pure $^4$He is larger 
than the one of the mixture at the same pressure and therefore the condensate
fraction in the mixture is larger than in $^4$He. However, as $P$ 
increases, the differences between the densities become smaller and the
condensate fractions of both systems get closer.

The low values of $Z_F$ imply a large value of the energy-dependent
effective mass at the Fermi surface,
\begin{equation}
M_E= 1 - {{\partial}\over {\partial E}}\Re \Sigma(p,E) \mid_{E=e_F,
p=p_F}{}= Z_F^{-1}
\end{equation}
where $\Sigma(p,E)$ is the self-energy of the $^3$He atoms in the
mixture. At $x_3=0.04$, $M_E= 12 m_3$ which is around three times larger
than for pure $^3$He at the saturation density. \cite{fabro92,moro97}
This large value of the energy-dependent effective mass can be
attributed to the correlations with the $^4$He atoms, and implies a small
value of the $k$ dependent effective mass in order to reproduce the total
effective mass that, at those small concentrations, can be taken
$m_{3}^{*}/m_3= 2.3$,\cite{Polls86,Edwards} i.e., 
the value in the impurity case.

Fig. 6 shows $n^{(4)}(k)/\rho_4$ and $\nu n^{(3)}/\rho_3$
 for a 6 $\%$ mixture (solid and long-dashed lines respectively) together 
with $n^{(4)}(k)/\rho_4$ for pure $^4$He at the equilibrium density 
(short-dashed). The three momentum distributions are very close above 
$k_F$, as the large-$k$ behavior is essentially dominated by the short-range 
dynamical correlations. 
As in the pure phases, the tails of the
momentum distributions ($k > 3.5$ \AA$^{-1}$) are taken to have an 
exponential behavior. Their contribution at $x=6.6 \%$ to the
total kinetic energy is $\sim 8 \%$.
On the other hand, the kinetic energy of the free Fermi sea (that
would give an upper-bound to the contribution to $T_3/N_3$ below $k_F$) 
is 0.58 K. That means that more than $97 \%$ of the $^3$He 
kinetic energy comes from momenta above $k_F$, clearly
showing the importance of the correlations between $^3$He and $^4$He
atoms.

It is also of interest to consider the dependence of $T_3/N_3$ on the 
concentration. Fig. 7 gives $T_3/N_3$ in function of the  
$^3$He partial density in the mixture along the $P=0$ isobar.
Obviously, the kinetic energy  ends up with the kinetic energy of
pure $^3$He  ($\sim 12$ K) which corresponds to a density value  
that lies out of the plot. Therefore the kinetic energy of the
$^3$He should be in average a decreasing function of the concentration
except for the behavior at the origin where the term associated with
the free Fermi kinetic energy dominates the overall decreasing behavior
driven by the decrease of the total density. Actually, the kinetic energy 
in the interval considered here is well parameterized as the sum of the 
free Fermi gas energy plus a linear term describing   
the decrease of the kinetic energy with the density

\begin{equation}
\frac{T_3}{N_3}= \frac{T_3}{N_3}(\rho_3=0) - A \rho_3 + \frac{3}{10}
\frac{\hbar^2}{m_3} \left( \frac{6 \pi^2}{\nu}\right)^{2/3} \rho_3^{2/3} \ .
\label{fitkin}
\end{equation}

The numerical value of the parameter $A$ may be
estimated by calculating the $x_3$ dependence of the kinetic energy in the
underlying boson-boson mixture and it results to be $A=27.2$ K$\sigma^3$.

\section{Discussion and conclusions}
The results obtained in this paper for the $^4$He condensate
fraction and the $x_3$ dependence of the $^3$He kinetic energy are in
contrast with recent experimental estimates. In fact, Sokol {\em et
al.},
\cite{sok94,sok95}, analyzing deep
inelastic neutron scattering measurements carried out for a $9.5 \%$
mixture at $1.4$ K, and for a momentum transfer as high as $23$ \AA$^{-1}$, 
estimated a condensate fraction $n_0^{(4)} = 18 \%$ and a $^3$He kinetic
energy of approximately $10$ K, basically independent on the
concentration. These results are to be compared with the theoretical
predictions $n_0^{(4)}\sim 10 \%$ and $T_3/N_3 \sim 19$ K obtained
in ACA for a  similar mixture.

It has been argued \cite{sok94} that the main source of
discrepancy with a preliminary presentation of the present results
\cite{boro90} is due to the use of ACA, implying  the same type
of local environment for the different types of atoms in the mixture.
Sokol's observation is physically founded on  the large zero
point motion of the $^3$He atoms which should decrease the local density
around them to a value similar to the pure $^3$He. Obviously, the
use of optimal correlations should clarify this point. However, it must be 
stressed that  the $T=0$ DMC calculations  of Ref. \onlinecite{boro96} give 
for the $^3$He impurity kinetic energy  $T_3= 17.5$ K, i.e. a $1.5 $ K 
lower value than the ACA prediction estimated by using the pure $^4$He 
DMC kinetic energy ($T_4=14.3$ K).\cite{boro94}
On the other hand, the predicted
$n_0^{(4)}$   by DMC \cite{boni96} points to
an extrapolated value of $11 \%$ for a 6.6 \% mixture at the same 
temperature. A dramatic change of both $n_0^{(4)}$ and  $T_3$ at higher 
concentrations would be required in order to reproduce the experimental 
estimates.

In conclusion, we believe that although the use of optimal correlations
will certainly decrease the kinetic energy of the $^3$He component and
enhance a little the $^4$He condensate fraction, the
resulting values will be far from the present experimental analysis. A full
theoretical calculation of the scattering process including final state
interactions and the experimental broadening, similar to the ones
performed in pure $^4$He,\cite{ferran96}
is necessary in order to fully understand the experimental measurements  
and reliably extract kinetic energies and condensate fractions..

Summarizing, we have calculated the momentum distributions of
$^3$He-$^4$He mixtures in the framework of the HNC/FHNC equations
using variational wave functions with two- and three-body correlations.
These momentum distributions can be used as input for the analysis of
the recent performed inelastic neutron scattering experiments. It has
been found that, at the low concentration where the mixture is stable, 
the Fermi statistics do not significantly modify the value of the $^4$He  
condensate
fraction. On the other hand, it is crucial to take into account the Fermi
statistics for the stability of the mixture.
The
concentration dependence of the different quantities studied in the
paper can be mainly explained by the decrease in the total density of the
mixture when the $^3$He concentration increases.

\acknowledgments
The authors are especially indebted with J. Casulleras for 
his collaboration in the $^4$He Monte Carlo calculations and with S. Moroni for 
several fruitful exchanges and discussions and for providing his 
VMC results.  This research was supported in part by DGICYT (Spain) Grant
No. PB95-0761, CICYT (Spain) Grant No. TIC95-0429, the agreement DGICYT
(Spain)--INFN (Italy) and the Accion Integrada Hispano-Italiana
99A-1994. 

\appendix     
\section*{}

 In this Appendix we present the HNC/FHNC equations for the mixture 
one-body density matrices.

The sums of the nodal diagrams contributions, $N_{\omega_c \omega_c}^{(3)}$ 
and $N_{\omega \omega}^{(\alpha )}$, are obtained by solving the integral 
equations 
\begin{equation}
N_{\omega \omega}^{(\alpha)}=\sum_{\lambda=3,4} \rho_{\lambda}
\sum_{z,y}
(g_{\omega z}^{(\alpha,\lambda)}-N_{\omega z}^{(\alpha,\lambda)}-\delta_{
zd} \mid g_{y\omega}^{(\lambda,\alpha)}-\delta_{yd}) \ ,
\label{nomega}
\end{equation}
and
\begin{eqnarray}
N_{\omega_c \omega_c}^{(3)} & = & \rho_3 (g_{\omega_c c}+l(k_F
r_{12})/{\nu}
- N_{\omega_c c}^{(3)} \mid g_{c \omega_c}+l/\nu)
\\ \nonumber
 & & +\rho_3 (-l/\nu\mid 2(g_{c\omega_c}+l
/\nu-
N_{c\omega_c}^{(3)})-(g_{cc}+l/\nu-N_{cc})) \ .
\label{nomc}
\end{eqnarray}
The notation $(A(r_{ij})\mid B(r_{jk}))$ stands for the convolution product

\begin{equation}
(A(r_{ij})\mid B(r_{jk}))=\int d{\bf r}_j A(r_{ij})B(r_{jk}) \ .
\label{convol}
\end{equation}
The summations over $z$ and $y$ (where $z,y=d,e,c$) always extend to
all possible connections allowed by the diagrammatic rules of the HNC/FHNC
theory.\cite{fabro182,boro90}

Besides the distribution functions $g_{zy}^{(\alpha,\beta)}(r)$ ($g_{dd}^{(
\alpha,\beta)},g_{de}^{(\alpha,3)},g_{ee}^{(3,3)}$ and $g_{cc}^{(3,3)}$),
which have been defined elsewhere,\cite{boro93,fabro82} it is necessary
to introduce the auxiliary distribution functions:
\begin{equation}
g_{\omega d}^{(\alpha,\beta)}(r)=f^{(\alpha,\beta)}(r) \exp[B_{\omega
d}^ {(\alpha,\beta)}(r)]  \  ,
\label{gwd}
\end{equation}
\begin{equation}
g_{\omega e}^{(\alpha,3)}(r)=g_{\omega d}^{(\alpha,3)}(r) B_{\omega e}^{(
\alpha,3)}(r)          \ ,
\label{gwe}
\end{equation}
\begin{equation}
g_{\omega_c c}^{(3,3)}(r)=g_{\omega d}^{(3,3)}(r)
\frac{L_{\omega}(r)}{ \nu}        \ ,
\label{gwcc}
\end{equation}
where
\begin{equation}
B_{\omega x}^{(\alpha,\beta)}(r)=N_{\omega x}^{(\alpha,\beta)}(r)+
E_{\omega x}^{(\alpha,\beta)}(r)+C_{\omega x}^{(\alpha,\beta)}(r)
\ ,
\label{bw}
\end{equation}
and
\begin{equation}
L_{\omega}(r)=-l(k_F r)+\nu B_{\omega_c c}^{(3,3)}(r) \ .
\label{lw}
\end{equation}
The functions $E_{\omega d}^{(\alpha,\beta)}(r),E_{\omega e}^{(\alpha,3)}(
r)$ and $E_{\omega_c c}^{(3,3)}(r)$ give the contributions of the elementary
diagrams.

The nodal functions $N_{\omega z}^{(\alpha,\beta)}(r)$ are 
solutions of the following integral equations:
\begin{equation}
N_{\omega x}^{(\alpha,\beta)}=\sum_{\lambda=3,4} \rho_\lambda \sum_{z,y}
(g_{\omega z}^
{(\alpha,\lambda)}-N_{\omega z}^{(\alpha,\lambda)}-\delta_{zd} \mid
g_{yx}^{(\lambda,\beta)}-\delta_{yd}) \ ,
\label{nomx}
\end{equation}
\begin{equation}
N_{\omega_c c}^{(3,3)}=\rho_3 (g_{\omega_c c} - N_{\omega_c c} + l/\nu 
\mid g_{cc}) \ .
\label{nomcx}
\end{equation}

Finally, the functions $C_{\omega x}^{(\alpha,\beta)}(r)$ give the
contribution of the $dressed$ triplet correlations,
\begin{equation}
C_{\omega x}^{(\alpha,\beta)}(r_{12})=\sum_{\lambda=3,4} \rho_{\lambda}
\int
d {\bf r}_3 \, \omega^{(\alpha,\lambda,\beta)}(r_{12},r_{13},r_{23})
\sum_{zy} g_{\omega z}^{(\alpha,\lambda)}(r_{13}) g_{yx}^{(\lambda,\beta)}
(r_{32}) \ ,
\label{comx}
\end{equation}
and
\begin{equation}
C_{\omega_c c}^{(3,3)}(r_{12})=\rho_3 \int d {\bf r}_3 \, 
\omega^{(3,3,3)}
(r_{12},r_{13},r_{23}) g_{\omega_c c}^{(3,3)}(r_{13}) g_{cc}^{(3,3)}(r_{32}
)   \ .
\label{comcx}
\end{equation}
The functions $N_{zy}^{(\alpha,\beta)}(r)$ and $C_{zy}^{(\alpha,\beta)}(r)$
have been defined in Ref. \onlinecite{boro93}.

The quantities
$\Gamma_{\omega}^{(\alpha)}$ and $\Gamma_d^{(\alpha)}$, entering the expressions
of the strength factors $n_0^{(\alpha)}$,  are given by
\begin{eqnarray}
\Gamma_x^{(\alpha)}& = & \sum_{\lambda=3,4} \rho_{\lambda} \int d {\bf r}
(g_{xd}^{(\alpha,\lambda)}(r)-1-N_{xd}^{(\alpha,\lambda)}(r)-
E_{xd}^{(\alpha,\lambda)}(r)) \nonumber \\
&  &  + \rho_3  \int d {\bf r}
(g_{xe}^{(\alpha,3)}(r)-N_{xe}^{(\alpha,3)}(r)- E_{xe}^{(\alpha,3)}(r)) 
\nonumber \\
&  &  - (1 / 2) \sum_{\lambda=3,4} \rho_{\lambda} \int d {\bf r}
(g_{xd}^{(\alpha,\lambda)}(r)- 1 + \delta_{\lambda 3} g_{xe}^{(\alpha,\lambda)}
(r)) (N_{xd}
^{(\alpha,\lambda)}(r)+ 2 E_{xd}^{(\alpha,\lambda)}(r)) \nonumber \\
&  &  - (1 / 2) \rho_3 \int d {\bf r}
(g_{xd}^{(\alpha,3)}(r)- 1 ) (N_{xe}
^{(\alpha,3)}(r)+ 2 E_{xe}^{(\alpha,3)}(r)) \nonumber \\
&  &  - (1 / 2) \sum_{\lambda=3,4} \rho_{\lambda} \int d {\bf r}
(g_{xd}^{(\alpha,\lambda)}(r) + \delta_{\lambda 3} g_{xe}^{(\alpha,\lambda)}(r)) C_{xd}
^{(\alpha,\lambda)}(r) \nonumber \\
&  &  - (1 / 2)  \rho_3 \int d {\bf r}
g_{xd}^{(\alpha,3)}(r)  C_{xe}^{(\alpha,3)}(r)  +  E_{x}^{(\alpha)}
\label{gama}
\end{eqnarray}
where $E_x^{(\alpha)}$ is the sum of the one-point elementary diagrams.
\cite{manou85,fabro92,boro90} By setting $\rho_3=0$ ($\rho_4=0$), expression
(2.15) reduces to the pure phases $\Gamma_x$.\cite{manou85,fabro92}

\begin{table}

\caption{$^4$He condensate fraction, $^3$He $Z_F$ factor and partial kinetic
energies in the mixtures as a function of the $^3$He concentration at zero
pressure. The first lines are the Jastrow values. The second lines include 
the effect of the triplet correlations.}

\begin{tabular}{cccccc}
$x_3$  & $\rho (\sigma^{-3})$ & $n_0$ & $Z$  &  $T_4/N_4 (K)$ & $T_3/N_3 (K)$
  \\ \tableline
0.0    & 0.3648   &  0.091   &          & 15.06 & 19.99    \\
       &          &  0.082   &          & 14.52 & 19.27    \\  \tableline
0.02   & 0.3629   &  0.092   & 0.093    & 14.92 & 20.04 \\
       &          &  0.085   & 0.085    & 14.39 & 19.33  \\  \tableline
0.04   & 0.3609   &  0.094   & 0.094    & 14.79 & 19.99  \\
       &          &  0.086   & 0.086    & 14.27 & 19.30  \\  \tableline
0.066  & 0.3582   &  0.096   & 0.096    & 14.61 & 19.88   \\
       &          &  0.088   & 0.088    & 14.10 & 19.21   \\
\end{tabular}

\end{table}

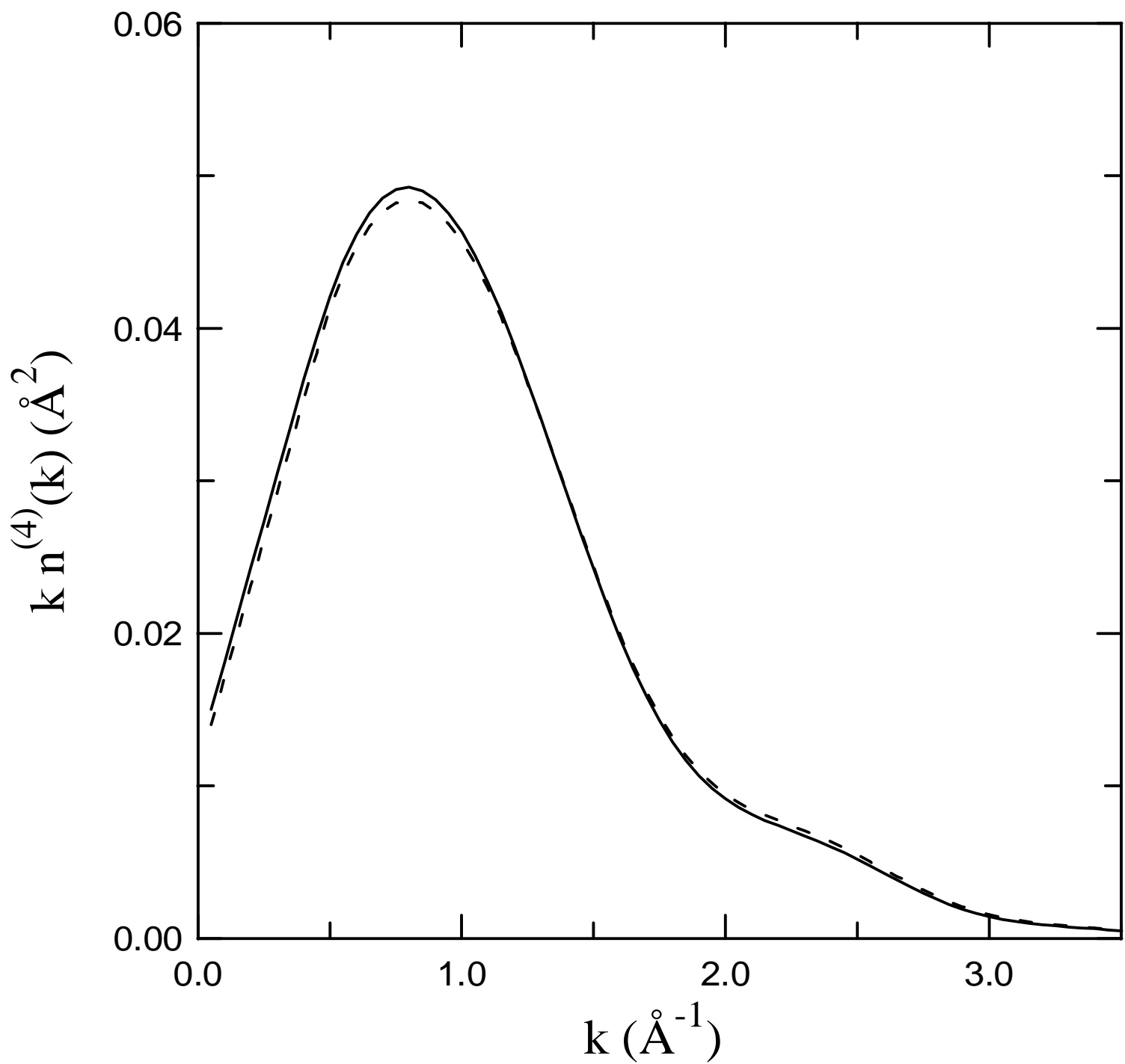
\begin{figure}
\caption{
Momentum distribution of the $^4$He atoms in the mixture. The continous
line corresponds to $x_3=0.066$ ($\rho =0.3582 \sigma^{-3}$) and the
dashed line to pure $^4$He at saturation density ($\rho=0.365
\sigma^{-3}$). Both results are at zero pressure.
}
\label{fig:nkhe4}
\end{figure}

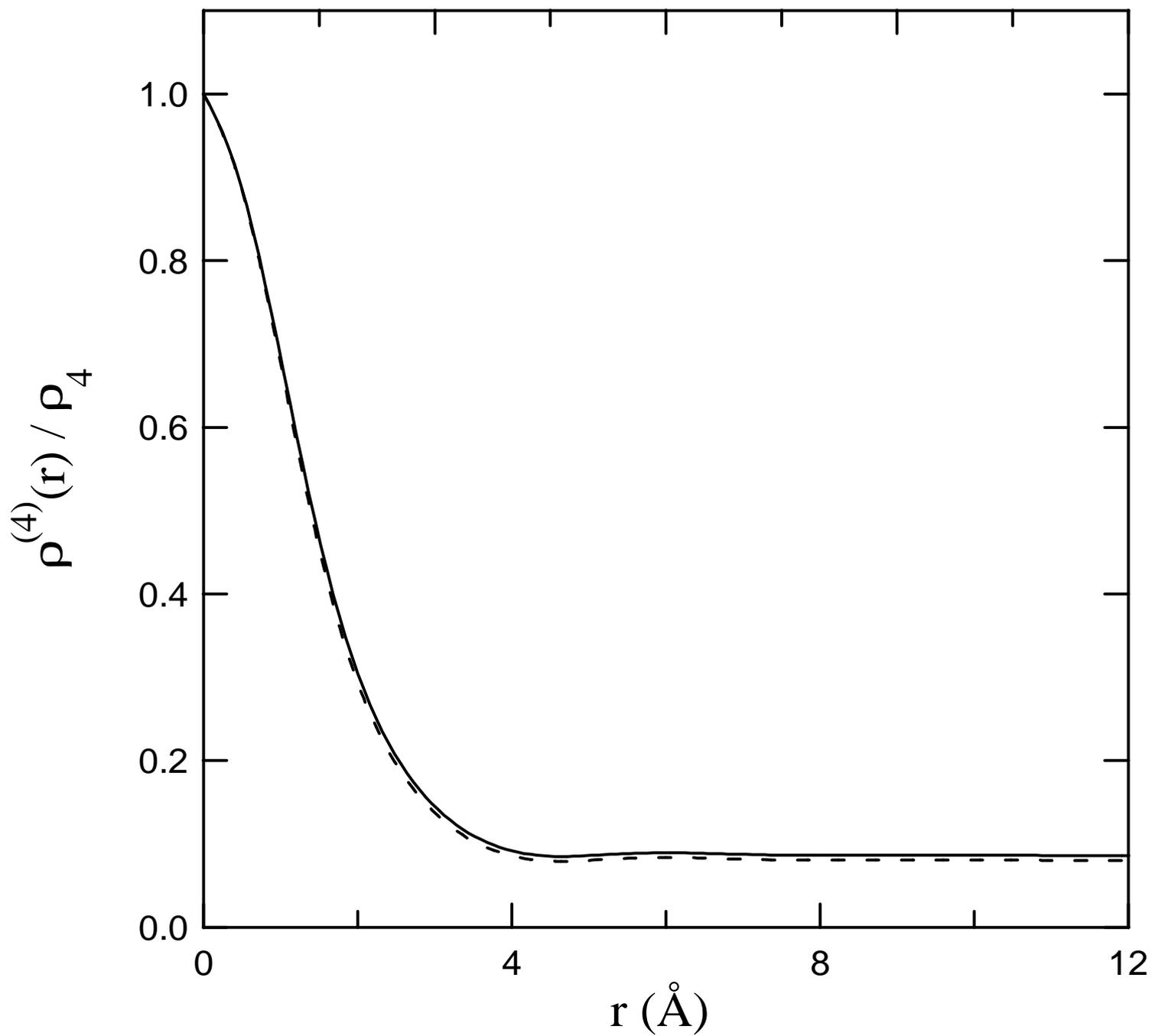
\begin{figure}
\caption{One-body density matrix of the $^4$He atoms in the mixture.
The notation is the same as in Fig. 1}
\label{fig:onhe4}
\end{figure}

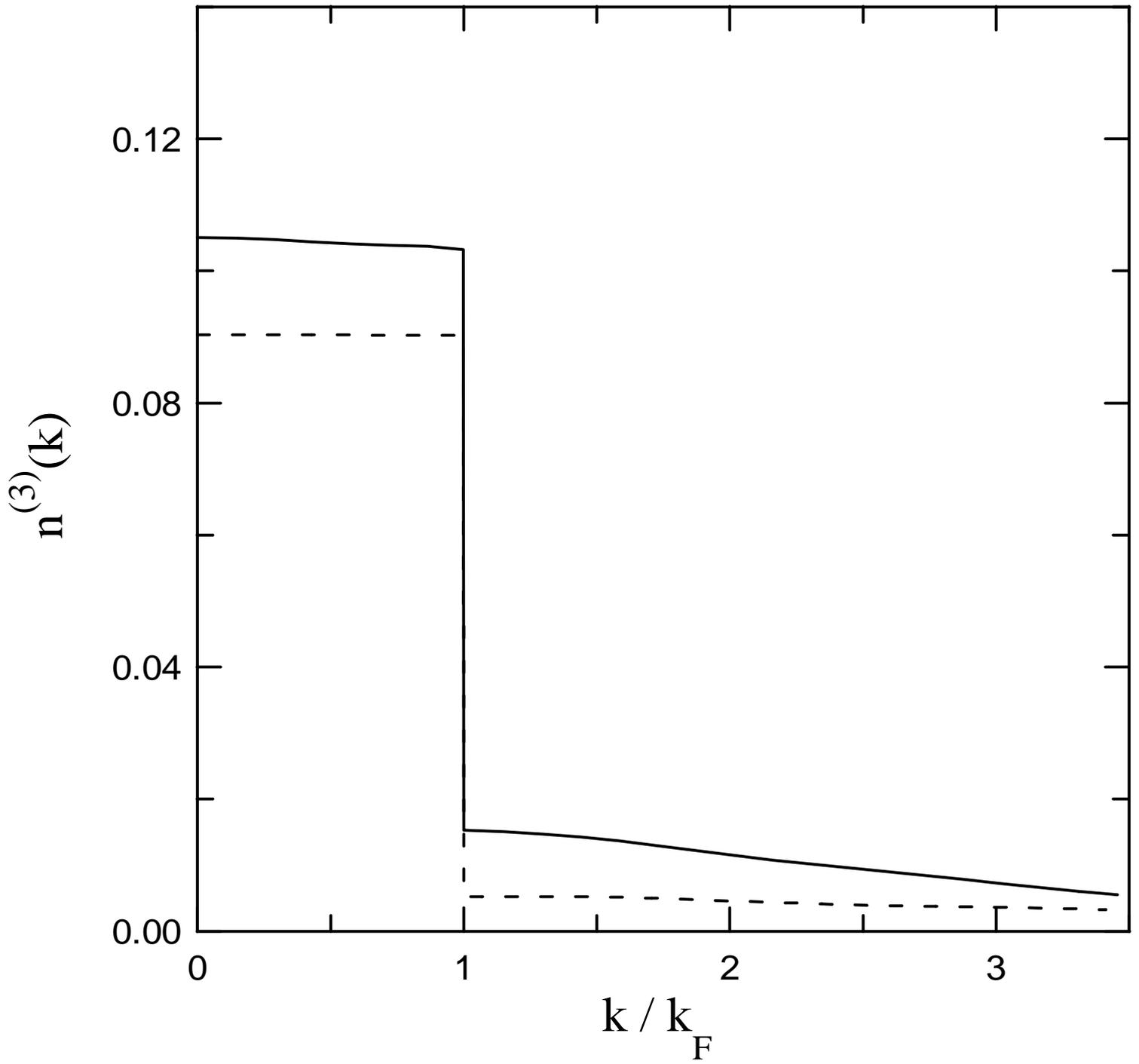
\begin{figure}
\caption{$^3$He momentum distributions in the mixture at $x_3=0.066$
(solid line) and $x_3=0.02$ (dashed line). The values of $k_F$ are 0.347
\AA$^{-1}$ and 0.235 \AA$^{-1}$, respectively.}
\label{fig:nkhe3}
\end{figure}

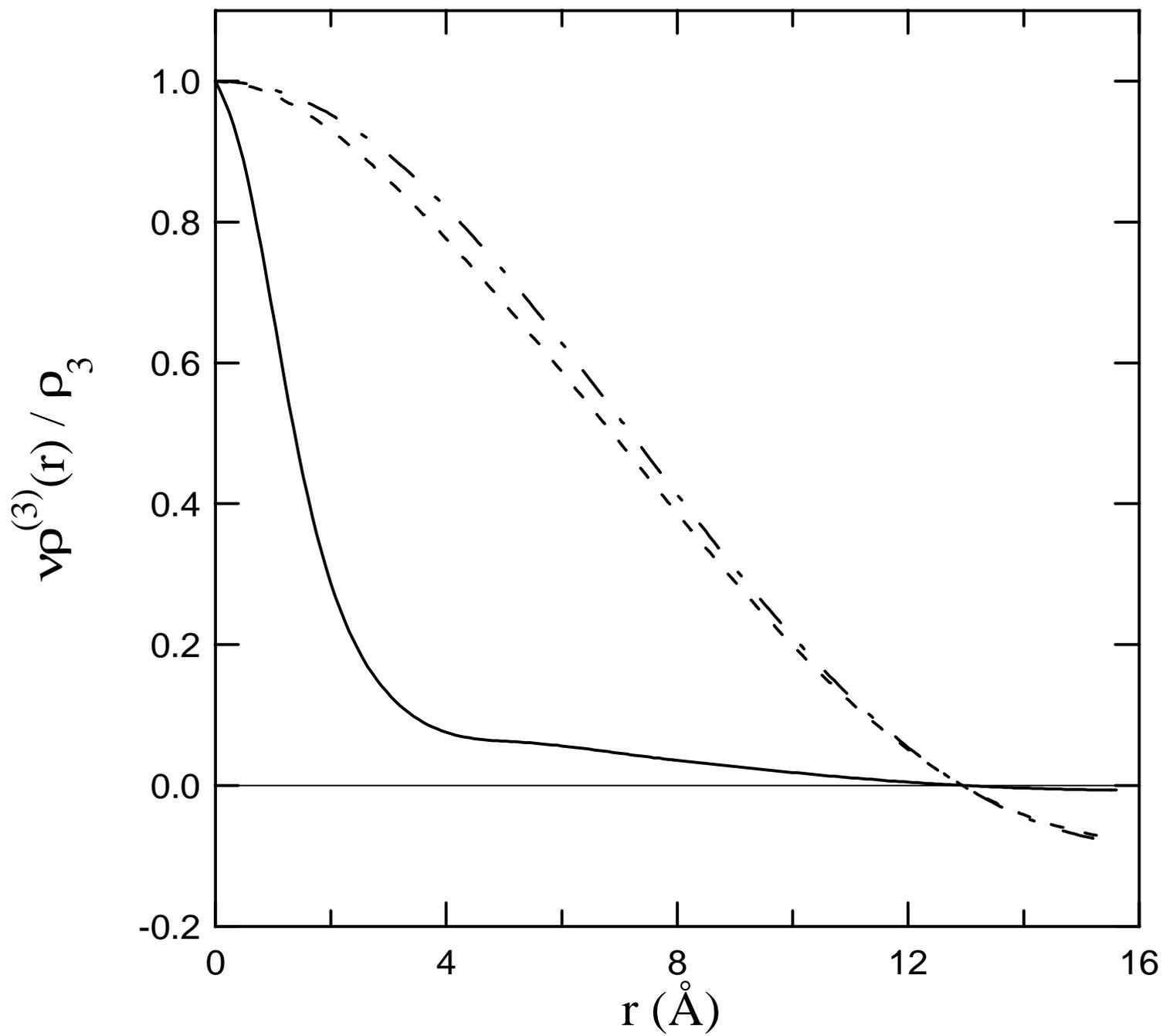
\begin{figure}
\caption{One-body density matrix of the $^3$He atoms in a $x_3=0.066$  mixture
(solid line) compared with the free Fermi system ( dash-dotted  line)
and pure $^3$He (dashed line), both at the same partial density $\rho_3$.}
\end{figure}

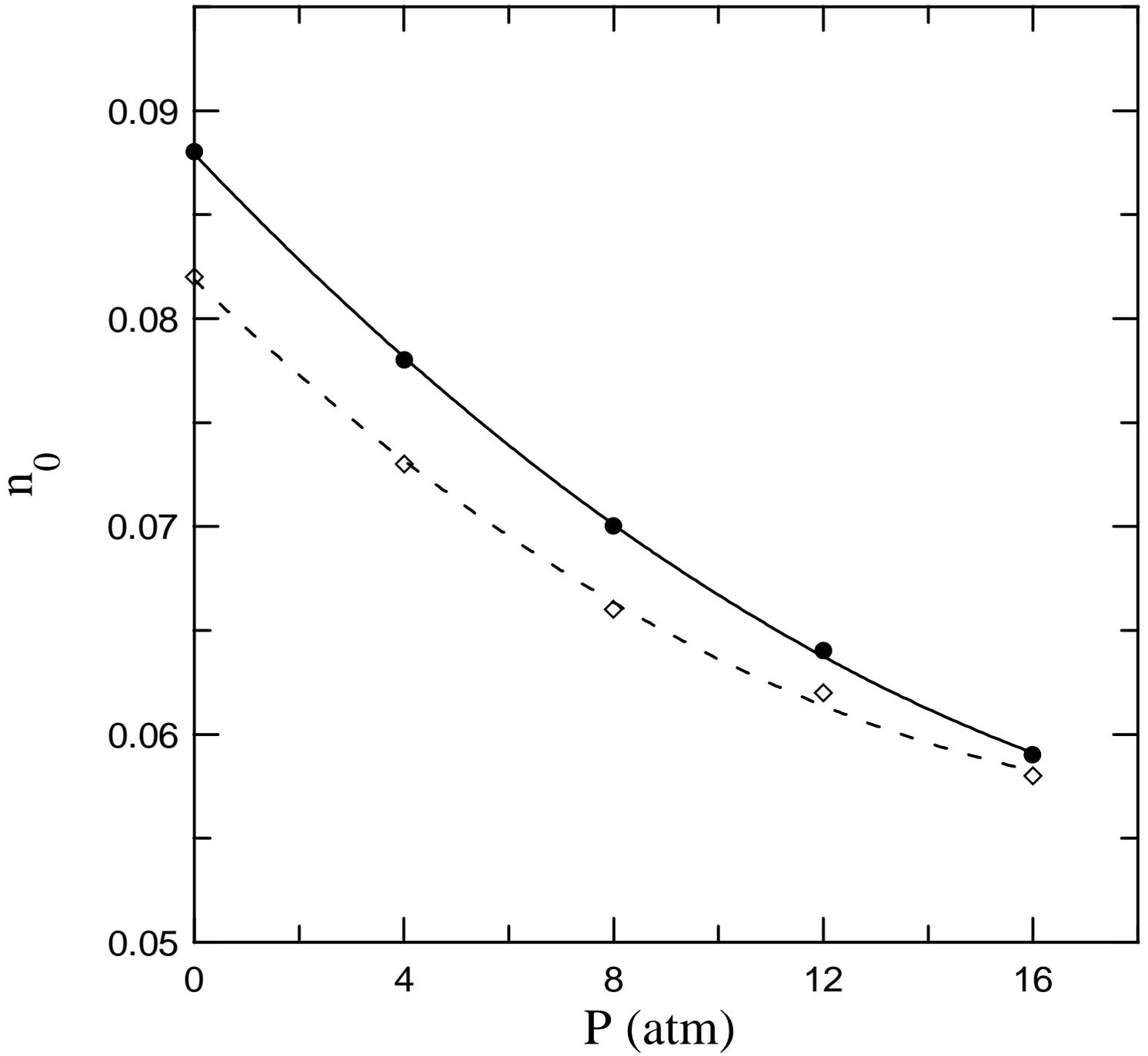
\begin{figure}
\caption{Condensate fraction as a function of pressure. The diamonds and
circles correspond to pure $^4$He and to a $x_3=0.066$ mixture, respectively. 
The lines are  guides to the eye.}
\end{figure}

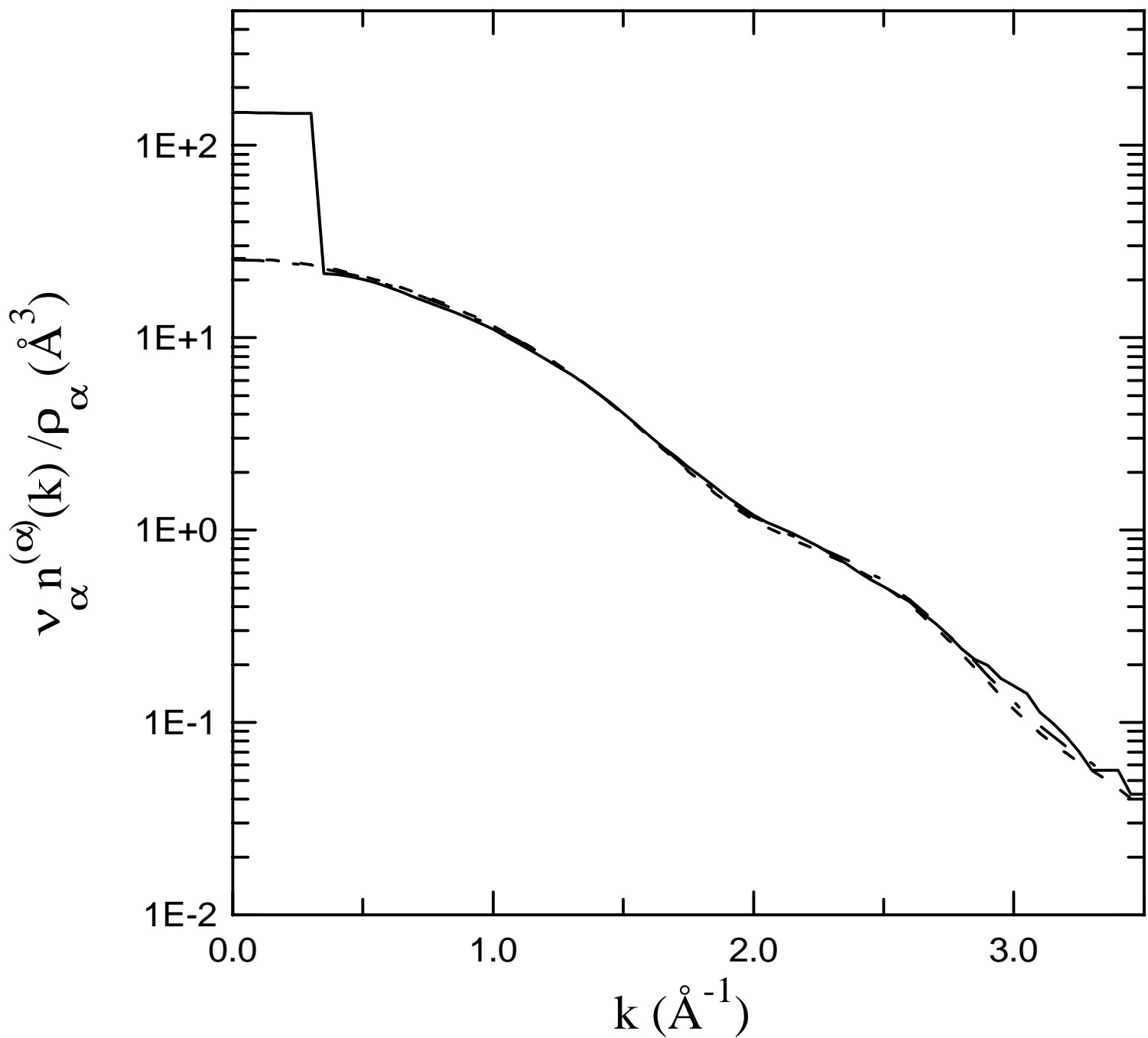
\begin{figure}
\caption{Momentum distributions per particle of pure $^4$He at equilibrium
density (short-dashed), and of $^4$He (long-dashed) and $^3$He (solid) of a
$x_3=0.066$ mixture.}
\end{figure}

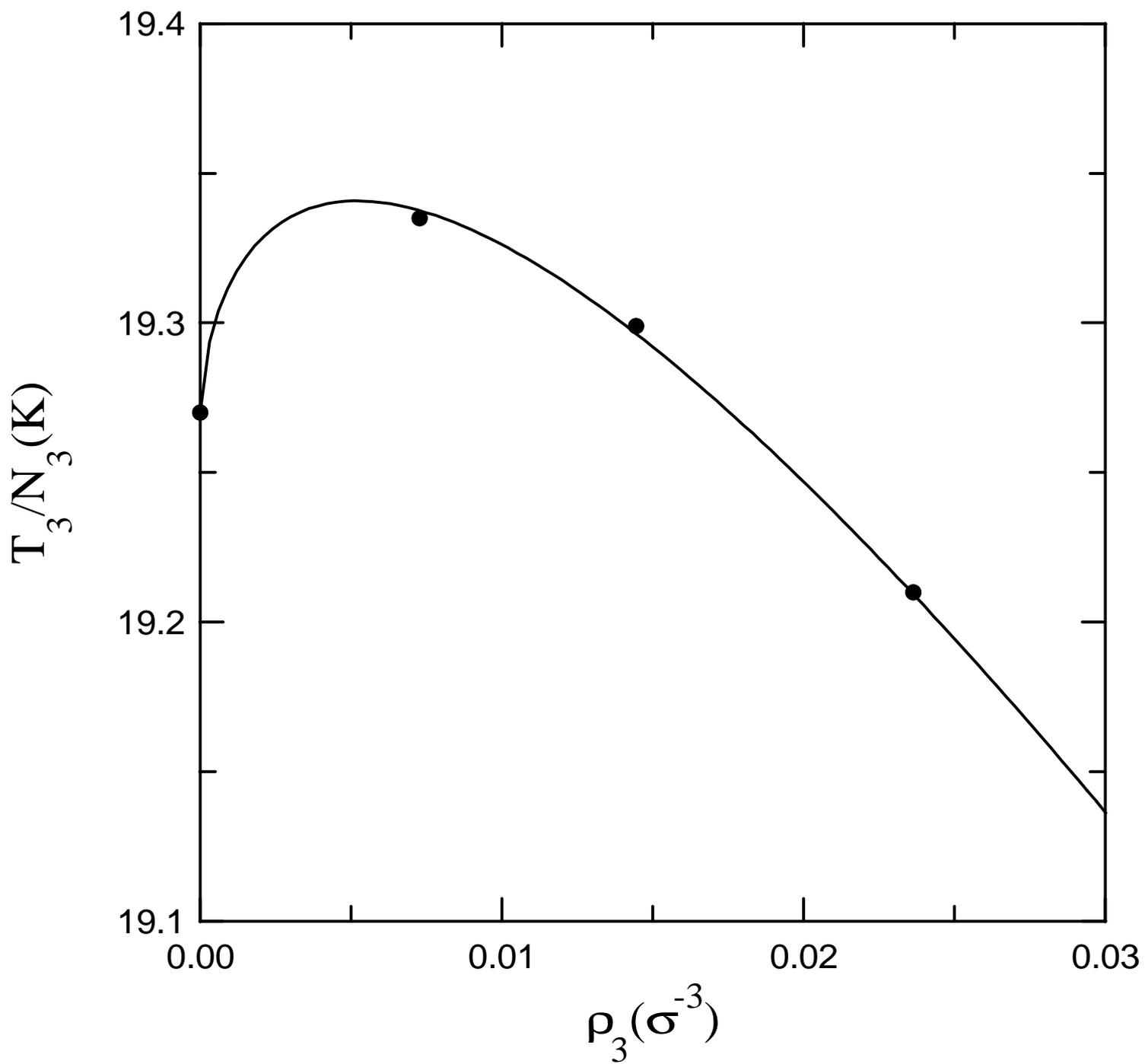
\begin{figure}
\caption{$^3$He kinetic energy as a function of $\rho_3$ at  
$P=0$. The solid line is the fit provided by Eq. (\protect\ref{fitkin}).}
\end{figure}

\end{document}